%% file: main.tex
\documentclass[journal,twoside,web]{ieeecolor}
\usepackage{generic}
\usepackage{amsmath,amssymb,amsfonts}
\usepackage{array}
\usepackage[caption=false,font=normalsize,labelfont=sf,textfont=sf]{subfig}
\usepackage{textcomp}
\usepackage{url}
\usepackage{verbatim}
\usepackage{graphicx}
\usepackage{cite}
\usepackage{float}
\usepackage[table]{xcolor}
\usepackage[T1]{fontenc}
\usepackage{CJKutf8}
\usepackage{tabularx}
\usepackage{booktabs}
\usepackage{multirow}
\usepackage{wrapfig}
\usepackage{arydshln}
\usepackage{threeparttable}
\usepackage{nicefrac}
\usepackage{microtype}      
\usepackage[utf8]{inputenc} 
\usepackage{hyperref}
\hypersetup{hidelinks=true}
\usepackage[ruled,linesnumbered,lined,boxed,commentsnumbered]{algorithm2e}

\def\BibTeX{{\rm B\kern-.05em{\sc i\kern-.025em b}\kern-.08em
    T\kern-.1667em\lower.7ex\hbox{E}\kern-.125emX}}

\begin{document}

\title{Physiological Prior-Driven Label Enhancement for Cross-Subject EEG Emotion Recognition}

\author{Hongyu Zhu, 
    Lin Chen,
    Yuming Fu,
    ~\IEEEmembership{Senior member, IEEE} Mounim~A.~El-Yacoubi,
    Mingsheng Shang
\thanks{H. Zhu, L. Chen, and M. Shang are with Chongqing Institute of Green Intelligent Technology, Chinese Academy of Sciences, e-mail: \{zhuhongyu, chenlin, msshang\}@cigit.ac.cn. (Corresponding
authors: Lin Chen; Mingsheng Shang)} 
\thanks{Y. Fu is with Central South University, School of Computer Science and Engineering, e-mail: fym291715@163.com.}
\thanks{M. A. El-Yacoubi is with SAMOVAR, Telecom SudParis, Institute Polytechnique de Paris, 91120 Palaiseau, France, e-mail: mounim.el\_yacoubi@telecom-sudparis.eu.}
\thanks{This work was supported in part by the Chongqing Key Project of Technological Innovation and Application (CSTB2023TIAD-STX0015, CSTB2025TIAD-STX0034) and the National Natural Science Foundation of China (Grant No. 62506054).}
}

\maketitle

\begin{abstract}
Electroencephalography (EEG)-based emotion recognition captures affective neural signals with high temporal precision, but cross-subject variability and label noise remain critical challenges to its practical healthcare deployment. Existing label-denoising methods lack physiological grounding, while physiology-informed approaches rely on hand-crafted hyperparameters. To bridge these two paradigms, we propose PhyDA, a plug-and-play, tuning-free framework that unifies neurophysiological priors with data-driven label refinement. PhyDA comprises two modules. {Since cross-subject variability renders global thresholds suboptimal,} the Physiological Noise Quantifier (PhyNQ) {exploits} {a spectral slope} to produce a subject-specific noise score, providing a neurophysiologically interpretable quality assessment {that naturally adapts to each individual}. The Data-Adaptive Label Refiner (DALR) directly adopts this noise score as the contamination ratio to drive a label refinement pipeline that requires no additional neural network training, thereby directly mitigating the impact of inter-subject label noise. Extensive experiments on three public datasets (DEAP, SEED, SEED-IV) across seven backbone architectures under strict leave-one-subject-out cross-validation demonstrate that PhyDA consistently and significantly outperforms both general and EEG-tailored label-denoising baselines, achieving average accuracy gains of 2.76\%, 2.66\%, and 3.32\%, respectively. Visualization further confirms its neurophysiological interpretability and practical robustness. The source code is available at: https://github.com/HongyuZhu-s/PhyDA.

\end{abstract}


\input{1_introduction}

\input{2_related}

\input{3_method}

\input{4_Implementation_Details}

\input{5_experiment}

\input{6_conclusion}

\bibliographystyle{IEEEtran}
\bibliography{reference}

\end{document}

%% file: 1_introduction.tex
\begin{figure*}[!ht]
\centerline{\includegraphics[scale=0.66]{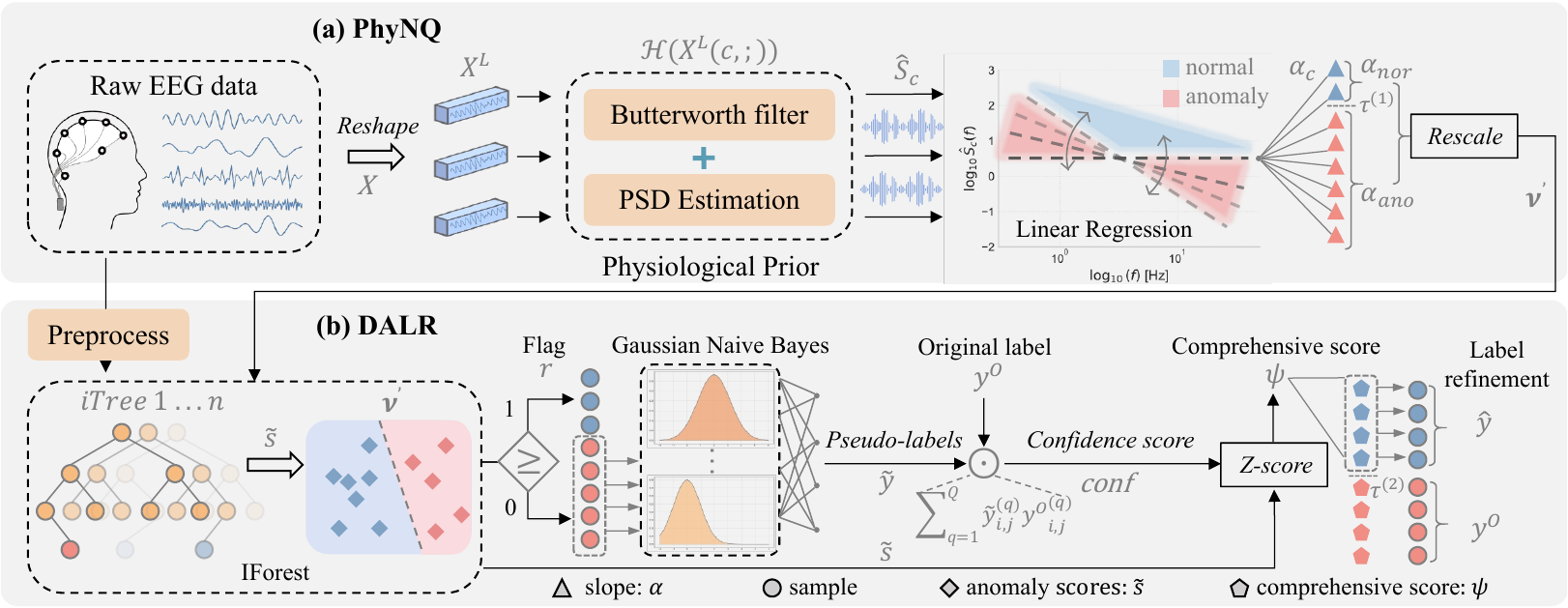}}
    \caption{Overview of the PhyDA framework.
    }
    \vspace{-5pt}
    \label{fig1}
\end{figure*}

\section{Introduction}
\IEEEPARstart{O}{wing} to its non-invasive nature and millisecond temporal resolution, electroencephalography (EEG) has become an important modality in brain-computer interfaces (BCI), clinical diagnostics, and cognitive neuroscience \cite{kakkos2021eeg}.
In affective analysis, EEG holds clear advantages over other modalities (e.g., facial expressions, speech, peripheral signals) due to its ability to capture neural correlates of emotion directly with high temporal precision. This has fueled rapid progress in EEG-based emotion recognition \cite{ding2022TSception,2017dreamer} and laid the groundwork for more intelligent and reliable human-computer interaction (HCI) \cite{mishra2025thought2textIT2, DingEEG}. 

Despite these advances, progress continues to be undermined by some inherent challenges: substantial inter-subject variability in EEG signals and label noise introduced by subjective emotional self-reporting \cite{hai2025DDSPR}, compounded by signal degradation from volume conduction effects \cite{shao2026exploringIT3}. {In typical EEG emotion benchmarks, labels range from binary to multi-class emotion categories, all derived from subjective post-stimulus ratings. The inherent ambiguity near decision boundaries and inter-individual differences in rating criteria make these labels particularly prone to noise, directly motivating the need for label refinement.} Together, these issues severely compromise the generalization of emotion recognition across subjects, hindering further research and practical deployment. Generalization across subjects is a fundamental challenge in applying deep learning to EEG decoding \cite{li2026LOSOsurvey, WangDBConformer}.

Existing strategies to alleviate label noise fall into three categories: sample selection \cite{han2018coteaching, yu2019doesCoteaching+}, label correction \cite{lidividemix2020, hai2025DDSPR}, and noise-robust loss functions \cite{hendrycks2018GLC, 2025real}. These methods typically require additional model training, thereby disregarding the supervisory information inherent in noisy data, and offer limited physiological interpretability.

In parallel, some works have introduced neurophysiological priors as auxiliary constraints to improve EEG signal modeling \cite{li2022eegRW1, donoghue2020RW5}. Although they improve representation, these priors almost invariably rely on manually crafted thresholds or dataset-specific hyperparameters, which compromise cross-subject generalization. Importantly, these two approaches have evolved in isolation from each other: existing label-denoising methods lack physiological grounding, and physiology-informed models have yet to be used for label refinement.

To overcome these limitations and enhance the model's generalization ability in cross-subject scenarios, we propose PhyDA, a tuning-free, subject-independent framework that adaptively enhances EEG label quality. 
{The core insight of this work is that because EEG emotion labels are obtained through subjective self-reports, label noise is subject-specific. Each subject's reporting bias creates a distinct noise pattern. A fixed threshold can never be optimal for all subjects. Physiological priors, such as the $1/f^\alpha$ spectral slope, are subject-specific by nature. They capture signal quality at the individual level without manual adjustments, making them ideal for driving per-subject adaptive label refinement.}
PhyDA consists of two modules, corresponding respectively to assessing data quality using neurophysiological priors and data-driven label refinement. The first module, Physiological Noise Quantifier (PhyNQ), exploits the $1/f^\alpha$ spectral slope as a principled, physiologically grounded indicator of signal quality. Without any manually specified thresholds, PhyNQ produces a normalized, subject-specific noise score $\nu_i \in [0, 1]$ that directly quantifies the noise pollution level of data. The second module is the Data-Adaptive Label Refiner (DALR). It uses the noise score as the contamination ratio to drive the label refinement process, which does not require any additional neural network training. Concretely, DALR first fits the Isolation Forest (IForest) \cite{liu2008isolation} {to obtain anomaly scores. The samples marked normal by IForest serve as a preliminary clean subset}, on which a lightweight Gaussian Naive Bayes (GNB) \cite{jahromi2017nonGNB} classifier is fitted to generate soft pseudo-labels for all samples. {Otsu’s method \cite{xu2011otsu} then adaptively partitions the samples based on a reliability score that combines the anomaly scores with pseudo-label confidence.} Only samples assigned to the noisy set undergo confidence-weighted label softening, while the original labels in the clean set are retained. Notably, every step is carried out independently per subject under strict leave-one-subject-out cross-validation (LOSO-CV) protocols.

The key insight is that the physiological noise score directly serves as the contamination ratio for label refinement, bridging two previously isolated paradigms. The main contributions of this work are as follows:

\begin{itemize}

\item We propose PhyDA, a plug-and-play label enhancement framework that bridges physiological modeling and data-driven label refinement. Without additional neural network training or hyperparameter tuning, PhyDA enhances cross-subject generalization of EEG-based emotion recognition.

\item We propose PhyNQ, a neurophysiologically grounded quality assessment module that exploits the $1/f^\alpha$ spectral slope to derive a subject-specific noise score, introducing physiological interpretability into the refinement pipeline.

\item We introduce DALR, a data-driven label refinement module that adopts the physiological prior as the contamination ratio, combining anomaly detection, pseudo-label generation, and adaptive label softening without discarding any sample.

\item Extensive experiments under strict LOSO-CV on three public datasets and seven backbones demonstrate that PhyDA significantly and consistently outperforms baselines, achieving average accuracy gains of 2.76\%, 2.66\%, and 3.32\% on DEAP, SEED, and SEED-IV, respectively.
\end{itemize}



%% file: 2_related.tex
\section{Related Work}
\subsection{Reduce Label Noise}
Label noise refers to the discrepancy between labels and the ground truth, commonly caused by annotation errors or subjective judgment biases, which leads supervised models to learn incorrect mappings \cite{frenay2013rw11}. 
To mitigate this issue, existing approaches fall into three categories. 
\paragraph{Sample‑level selection} A representative strategy is the small-loss sample selection. Co‑teaching \cite{han2018coteaching}, CoR \cite{wei2020CoR}, and Co-teaching+ \cite{yu2019doesCoteaching+} use the small-loss samples as clean instances and input them into an additional network for training as supervision. Upon this, Co‑learning \cite{2021Colearning} trains a network with structural similarity regularization and retains only the small‑loss samples for denoising.
Similarly, Cheng et al. \cite{chenglearningWR12} propose a confidence‑based selection criterion, while Toneva et al. \cite{tonevaempiricalWR13} track forgetting events across training epochs and treat rarely forgotten examples as reliable. However, these strategies typically require multiple epochs of model training and ignore the potential supervisory value of remaining samples.

\paragraph{Robust loss function}
Some methods improve the model's generalization by modifying the loss function, thereby mitigating the risk of overfitting due to label noise. For instance, Symmetric Cross Entropy loss \cite{wang2019symmetricCE} addresses the issue of gradient explosion and feature distortion caused by noisy labels, by combining the Cross Entropy loss with a Reverse Cross Entropy loss term. Another robust loss strategy, Gold Loss Correction \cite{hendrycks2018GLC}, uses trusted examples to efficiently compute the matrix and adjust the training loss. Recently, REAL \cite{2025real} has been proposed specifically for EEG-based classification, which integrates contrastive learning for shared representation extraction and iterative low-confidence sample filtering to mitigate label noise. However, in practical applications, relying on modifying the loss function often encounters problems such as slow model convergence, unstable training, and sensitivity to noise types.

\paragraph{Label correction}
Several studies transform the task of learning with noisy labels into semi-supervised learning, using pseudo-labels instead of noisy labels. DivideMix \cite{lidividemix2020} separates clean and noisy samples via a loss-based Gaussian Mixture Model and treats the noisy subset as unlabeled data for semi-supervised refinement. Building on this direction, Robust LR \cite{chen2023RobustLR} addresses confirmation bias in the two-stage framework by integrating pseudo-labeling with confidence estimation to refurbish noisy labels without discarding them. More recently, in the EEG domain, DDSPR \cite{hai2025DDSPR} mitigates label noise by dynamically filtering source domains and refining low-confidence pseudo-labels in the target domain.

Our proposed PhyDA framework is classified as a label correction method. Its key feature is that it improves interpretability by incorporating physiological prior knowledge, without requiring any model training or hyperparameter tuning.

\subsection{Physiological Modeling in EEG Data}
Physiological modeling has become an effective approach to improve the robustness of EEG analysis by incorporating neurophysiological priors \cite{li2022eegRW1}. In EEG-based affective computing, physiological modeling has predominantly focused on extracting features or designing constraints that reflect underlying neural processes. Unlike data-driven methods, it leverages intrinsic neural characteristics to guide signal processing and quality assessment. The most common practices rely on frequency-band power and asymmetry indices between symmetrical electrodes \cite{li2022eegRW1, hu2019tenRW2} or on phase synchronization measures \cite{zhu2026bimoeRW3}. However, their reliance on manually specified frequency bands, electrode groupings, or baseline intervals introduces dataset-dependent hyperparameters that undermine cross-subject generalization.

Unlike the above, the $1/f^\alpha$ power-law decay of the EEG spectrum is a more fundamental physiological property, and an important measurement of underlying neural activity \cite{donoghue2020RW5}. This aperiodic scaling reflects neural dynamics and is highly sensitive to noise, arousal, and individual differences \cite{ahmad20221fRW4}. Previous studies have used the $1/f^\alpha$ spectral slope in their research, such as Gao et al.\cite{gao2017RW6}, which uses the $1/f^\alpha$ exponent as an estimator for the excitation-inhibition (E-I) balance; The states of sleep \cite{lendner2020RW7} and anesthesia \cite{waschke2021RW8} are distinguished through $1/f^\alpha$ exponent.
Recent studies have found that high-quality EEG data should exhibit a reasonable $1/f^\alpha$ slope \cite{fasol20231fspectral, kozhemiako2022RW9}. Therefore, this slope can be used as an indicator of noise pollution to evaluate the quality of EEG data \cite{troendle2026rw10}. 

Nevertheless, existing methods for quality evaluation based on $1/f^\alpha$ spectrum still rely on hand-crafted thresholds to define abnormal data, neglecting substantial inter-subject variability in EEG signals. Furthermore, these methods have predominantly been validated on datasets with well-separated classes and low within-class variability, such as sleep staging. When applied to sentiment analysis datasets, where the data distribution is more complex, a purely threshold-based metric often fails to identify anomalous samples effectively.

%% file: 3_method.tex
\section{Methodology}

\subsection{Problem Setup}
Let $\mathcal{D} = \{\mathcal{D}_i\}_{i=1}^M$ denote a multi-subject EEG dataset comprising $M$ subjects, where $\mathcal{D}_i = (\mathbf{X}_i, \mathbf{Y}_i)$ represents the dataset of subject $i$. The raw signals are given by $\mathbf{X} \in \mathbb{R}^{S \times K \times C \times T}$, where $S, K, C, T$ denote the number of sessions, trials, electrode channels, and time points, respectively. The original labels $\mathbf{Y} \in \mathbb{R}^{S \times K}$ indicate the emotional states.
After preprocessing, the data is denoted $\mathbf{Z} \in \mathbb{R}^{N \times C \times D}$, where $N$ is the number of samples and $D$ is the feature dimension. 
The corresponding labels $\mathbf{Y}^{O} \in \mathbb{R}^{N \times Q}$ are converted into one-hot form, where $Q$ is the number of emotion classes.

The usual supervised objective is to learn a mapping $\mathcal{F}: \mathbf{X} \mapsto \mathbf{Y}$ that minimizes empirical risk under the original hard labels during training. Under this work, the training labels $\mathbf{Y}$ are refined into an enhanced set $\mathbf{Y}'$ to mitigate the adverse effects of label noise. Consequently, our goal is to determine an optimal mapping $\mathcal{F}': \mathbf{X} \mapsto \mathbf{Y'}$, thereby getting more robust and generalizable representations across subjects.

\input{tabs/Algorithm}

\subsection{Framework Overview}
An overview of PhyDA is shown in Fig.~\ref{fig1}. PhyDA comprises two modules. (1) PhyNQ exploits the intrinsic $1/f^\alpha$ spectral slope to produce a subject-specific noise-level estimate $\textbf{v}=[\nu_1,\nu_2,\dots,\nu_M]$ with $\nu_i \in [0,1]$, independently for each individual. (2) DALR adopts $\textbf{v}$ as the contamination ratio to adaptively partition each subject's samples into a clean set $\mathcal{D}^{clean}$ and a noisy set $\mathcal{D}^{noisy}$. Samples in $\mathcal{D}^{clean}$ retain their original labels, while those in $\mathcal{D}^{noisy}$ undergo confidence-weighted label softening. The complete process is presented in Algorithm \ref{alg:phyda}.

\subsection{Physiological Noise Quantifier}
In the PhyNQ module, we perform cross-subject noise quantification based on physiological constraints.
For a given LOSO-CV training fold, the raw EEG data $\mathbf{X}$ are first reshaped into $\mathbf{X}^{\text{L}} \in \mathbb{R}^{C \times (S \cdot K \cdot T)}$ by concatenating all sessions and trials along the temporal dimension. 

To filter out high and low frequency noise \cite{hu2025strflnet}, a 4th-order Butterworth \cite{shouran2020butterworth} band-pass filter $\mathcal{H}(\cdot)$ with a cutoff frequency of 0.3–50 Hz is applied to each channel signal $\mathbf{x}_c = \mathbf{X}^{\text{L}}(c,:)$, $c = 1, \dots, C$.
The EEG power spectrum is known to follow a $1/f^\alpha$ decay in the log‑log domain, where the spectral slope $\alpha$ reflects the scale‑free dynamics of neural activity and is sensitive to noise contamination \cite{fasol20231fspectral}.
Let $f_k$ denote the frequency at the $k$-th bin of the power spectral density (PSD) estimate $\hat{S}_c$ of the filtered signal $\mathcal{H}(\mathbf{x}_c)$.
For each channel $c$, the spectral slope $\alpha_c$ is obtained by ordinary least-squares linear regression on $\log_{10}\hat{S}_c(f_k)$ over the frequency band of interest (0.3--50\,Hz) 
Specifically, the slope $\alpha_c$ is estimated as:

\begin{equation}
    \begin{aligned}
    \label{eq1}
        \alpha_c = \arg\min_{\alpha, \beta} \sum_{k \in \mathcal{K}} \Bigl( \log_{10}\hat{S}_c(f_k) - \bigl( \beta - \alpha \log_{10} f_k \bigr) \Bigr)^2,
    \end{aligned}
\end{equation}
where $\mathcal{K} = \{ k \mid 0.3 \le f_k \le 50~\text{Hz} \}$, $\beta$ is the intercept.

To avoid manual thresholding, for each subject, we apply the Otsu method \cite{xu2011otsu} to the set of channel slopes $\{\alpha_c\}_{c=1}^C$. The algorithm automatically partitions the slopes into normal and anomalous categories by maximizing the inter-class variance, yielding an adaptive threshold $\tau^{(1)}$. Let $\alpha_{\text{nor}}$ and $\alpha_{\text{ano}}$
be the mean slopes of the normal and anomalous groups, which are separated by $\tau^{(1)}$. These points serve as subject-specific references for each channel.
For each channel with estimated slope $\alpha_c$, we compute a noise score based on $\alpha_{\text{nor}}$ and  $\alpha_{\text{ano}}$, using linear rescaling.  The result is clipped to $[0,1]$ to prevent extreme values from exceeding the unit interval.

\begin{equation}
    \begin{aligned}
    \label{eq2}
        \nu_c = \text{clip}(\frac{\alpha_{\text{nor}}-\alpha_c}{\alpha_{\text{nor}}-\alpha_{\text{ano}} + \epsilon}, \; 0, \; 1),
    \end{aligned}
\end{equation}
where $\epsilon$ is a minimum value, preventing the denominator from being zero. Finally, the physiological noise $\nu_i$ for subject $i$ is obtained by averaging across all channels:

\begin{equation}
    \begin{aligned}
        \nu_i= \frac{1}{C} \sum_{c=1}^C \nu_c.
    \end{aligned}
\end{equation}
This procedure is performed independently for every training subject within the current LOSO-CV fold, producing a noise vector $\textbf{v}'= [\nu_1, \nu_2, \dots, \nu_{M-1}]^\top \in \mathbb{R}^{M-1} $ that serves as the subject-independent physiological prior.

\subsection{Data-Adaptive Label Refiner}
For each training subject within a given LOSO-CV fold, the DALR module uses the noise prior $\nu_i$ from PhyNQ to guide label refinement adaptively. {DALR employs two partitions with distinct purposes: a preliminary IForest-based split that supplies a clean-enough training set for GNB, and a final Otsu-based split that determines which samples receive label softening.} First, we derive a contamination estimate through IForest \cite{liu2008isolation}. Then, we compute pseudo-labels via a GNB classifier \cite{jahromi2017nonGNB} fitted on the preliminary clean subset. Finally, we use Otsu \cite{xu2011otsu} to divide the dataset into $\mathcal{D}^{clean}$ and $\mathcal{D}^{noisy}$, and perform adaptive label softening on $\mathcal{D}^{noisy}$. This process is carried out without training or hyperparameter tuning, while strictly preserving subject-wise isolation.

DALR first uses an IForest detector to fit $\mathbf{Z}_i$ of each training subject $i$, producing a vector of \textit{anomaly scores} $\mathbf{s}_i$. The raw score for the $j$‑th sample $\mathbf{z}_{i,j}$ is defined by the standard isolation depth expectation:
\begin{equation}
    \begin{aligned}
        s_{i,j}=2^{-\mathbb{E}[h(\mathbf{z}_{i,j})] / {c(n_i)}},
    \end{aligned}
\end{equation}
where $h(\mathbf{z}_{i,j})$ is the path length of $\mathbf{z}_{i,j}$ in an isolation tree, $n_i$ is the effective subsample size, which is set to 512 by default, and $c(\cdot)$ is the normalizing constant for a binary search tree.

To eliminate inter‑subject scale variations and map the scores to a common $[0,1]$ range, we apply min‑max normalization:
\begin{equation}
    \begin{aligned}
        \tilde{s}_{i,j} = \frac{s_{i,j} - \min(\mathbf{s}_i)}{\max(\mathbf{s}_i) - \min(\mathbf{s}_i)}.
    \end{aligned}
\end{equation}
Then, the contamination ratio is directly taken from the
physiological prior $\nu_i$, so that no additional parameters are introduced. The anomaly flag $r_{i,j}$ is obtained as Eq.~\eqref{eq6}. Here, 1 denotes an anomalous sample, and 0 denotes a normal sample.
\begin{equation}
    \begin{aligned}
        \label{eq6}
        r_{i,j} = \mathbb{I}\Big[ \tilde{s}_{i,j} \ge Q_{1-\nu_i}(\tilde{\mathbf{s}}_i) \Big],
    \end{aligned}
\end{equation}
where $\mathbb{I}[\cdot]$ is the indicator function. $Q_{\delta}(\cdot)$ denotes the empirical $\delta$-quantile, returning the threshold whose proportion in the given set is exactly $\delta$.

We employ a GNB classifier to generate pseudo-labels for all samples due to its computational efficiency, well-calibrated probabilistic output and robustness when trained on small datasets. The GNB estimates the posterior probability based on the prior probability and likelihood. Concretely, for subject $i$, GNB is fitted exclusively on the normal samples $\mathbf{Z}^{\mathcal{C}}$ and their corresponding labels $\mathbf{Y}^{O,\mathcal{C}}$ identified by the anomaly detector:

\begin{equation}
    \begin{aligned}
        \mathcal{C}_i = \{\, j \mid r_{i,j} = 0 \,\}, \quad
        \mathbf{Y}_i^{O,\mathcal{C}} = \{y_{i,j}^O \mid j \in \mathcal{C}_i\}.
    \end{aligned}    
\end{equation}

For each class $q \in \{1, \dots, Q\}$, the class prior is estimated from the normal labels.

Specifically, Gaussian parameters from $\mathbf{Z}_i^{\mathcal{C}}$ and $\mathbf{Y}_i^{O,\mathcal{C}}$, are calculated as follows:
\begin{equation}
    \begin{aligned}
        P(q) = \frac{|\mathcal{C}_{i,q}|}{|\mathcal{C}_i|}, \qquad
        \mathcal{C}_{i,q} = \{\, j \in \mathcal{C}_i \mid \arg\max \mathbf{y}_{i,j}^O = q \,\}.
    \end{aligned}
\end{equation}
Then, the per‑feature mean $\boldsymbol{\mu}$ and variance $\boldsymbol{\sigma}^2$ for class $q$ are computed on 
$\mathbf{Z}_i^{\mathcal{C}}$:
\begin{equation}
    \begin{aligned}
        \boldsymbol{\mu}_{q} = \frac{1}{|\mathcal{C}_{i,q}|} \sum_{j \in \mathcal{C}_{i,q}} \mathbf{z}_{i,j},
        \quad
        \boldsymbol{\sigma}_{q}^2 = \frac{1}{|\mathcal{C}_{i,q}|} \sum_{j \in \mathcal{C}_{i,q}} 
        (\mathbf{z}_{i,j} - \boldsymbol{\mu}_{q})^2,
    \end{aligned}
\end{equation}
where the square and division are element‑wise.

After fitting, GNB computes soft pseudo‑labels for $j$-th sample of subject $i$, 
including both normal and anomalous ones, as the posterior probability vector:
\begin{equation}
    \begin{aligned}
    \label{eq:gnb_soft}
        \tilde{\mathbf{y}}_{i,j} = 
        \left( 
        \frac{P(q) \prod_{d=1}^{D} \mathcal{N}(\mathbf{z}_{i,j,d} \mid \mu_{q,d}, \sigma_{q,d}^2)}
        {\sum_{q'=1}^{Q} P(q') \prod_{d=1}^{D} \mathcal{N}(\mathbf{z}_{i,j,d} \mid \mu_{q',d}, \sigma_{q',d}^2)} 
        \right)_{q=1}^{Q},
    \end{aligned}
\end{equation}
where $\mathcal{N}(\cdot \mid \mu, \sigma^2)$ denotes the probability density function of a univariate Gaussian distribution, $\mathbf{z}_{i,j,d}$ is the $d$-th component of the feature vector of the $j$-th sample from subject $i$.

After obtaining the soft pseudo-labels $\tilde{\mathbf{y}}_{i,j}$, we compute the per‑sample \textit{confidence score}, which measures the agreement between the original label and the predicted distribution:
\begin{equation}
    \begin{aligned}
        {conf}_{i,j} = \sum_{q=1}^{Q} \tilde{y}_{i,j}^{(q)} \, y_{i,j}^{O(q)}.
    \end{aligned}
\end{equation}

To comprehensively evaluate the reliability of the labels, we construct a \textit{comprehensive score} $\psi$ by combining the scores of \textit{anomaly} and \textit{confidence}. Although manually tuning or iteratively optimizing the weights may improve performance on specific datasets, we opt for equal-weight summation. Both scores are Z-score standardized independently, making equal weighting a natural choice that preserves generality without requiring additional training.
$\psi_{i,j}$ is obtained by summing $\tilde{s}_{i,j}$ and $1 - {conf}_{i,j}$ with equal weight, after each variable has been independently standardized using Z-scores:
\begin{equation} 
    \begin{aligned}
    \psi_{i,j} = \frac{\tilde{s}_{i,j} - \mu_{\mathbf{s}_i}}{\sigma_{\mathbf{s}_i} + \epsilon}+ \frac{(1 - {conf}_{i,j}) - \mu_{\mathbf{conf}_i}}{\sigma_{\mathbf{conf}_i} + \epsilon}.
    \end{aligned}
\end{equation}

For subject $i$, we further employed Otsu method on $\mathbf{\psi}_i$ to automatically determine the optimal threshold $\tau^{(2)}_i$ that maximizes the between-class variance. 
This yields a subject-adaptive partition:
\begin{equation}
    \begin{aligned}
       \mathcal{D}^{noisy}_i = \{\, j \mid \psi_{i,j} > \tau^{(2)}_i \,\}, \\
        \mathcal{D}^{clean}_i = \{\, j \mid \psi_{i,j} \le \tau^{(2)}_i \,\}.
    \end{aligned}
\end{equation}    
Samples in $\mathcal{D}^{clean}_i$ retain their original labels $\mathbf{Y}_{i}^{O}$, and in $\mathcal{D}^{noisy}_i$, we perform adaptive label softening:
\begin{equation}
    \hat{\mathbf{y}}_{i,j} = {conf}_{i,j} \cdot \mathbf{y}_{i,j}^{O} + (1 - {conf}_{i,j}) \cdot \tilde{\mathbf{y}}_{i,j}.
\end{equation}
Thus, the refined label set $\mathbf{Y}'$ is defined as the union of
subject‑wise refined labels:
\begin{equation}
\mathbf{Y}' = \bigcup_{i \in \mathcal{I}_{\text{train}}} \mathbf{Y}'_i,
\end{equation}
where $\mathcal{I}_{\text{train}}$ denotes the set of training subjects in the
current LOSO‑CV fold, and for subject $i$,
\begin{equation}
\mathbf{Y}'_{i,j} =
\begin{cases}
\mathbf{y}_{i,j}^{O}, & j \in \mathcal{D}^{\text{clean}}_i,\\[4pt]
\hat{\mathbf{y}}_{i,j}, & j \in \mathcal{D}^{\text{noisy}}_i.
\end{cases}
\end{equation}

\subsection{Training with Refined Labels}
Let $\mathcal{D}_{\text{train}} = \{ (\mathbf{z}_{i,j}, \mathbf{y}'_{i,j}) \mid
i \in \mathcal{I}_{\text{train}},\; j = 1,\dots,N_i \}$ be the collection of all
training samples with their refined labels.
The emotion classifier $f_\theta$ is then trained by minimizing the standard
cross‑entropy loss:
\begin{equation}
\mathcal{L}_{\text{cls}} =
\frac{1}{|\mathcal{D}_{\text{train}}|}
\sum_{(\mathbf{z}, \mathbf{y}') \in \mathcal{D}_{\text{train}}}
\sum_{q=1}^{Q} -\, y'^{(q)} \log p_\theta^{(q)}(\mathbf{z}),
\end{equation}
where $p_\theta$ denotes the classifier's softmax output.

%% file: tabs/Algorithm.tex
\begin{algorithm}[!ht]
\caption{PhyDA}
\label{alg:phyda}
\KwIn{Raw EEG data $\mathbf{X}$; preprocessed features $\mathbf{Z}$ and labels $\mathbf{Y}^{O}$}
\KwOut{Refined training labels $\mathbf{Y}'$}
\BlankLine

\For{each LOSO fold}{
    \tcp*[h]{\textbf{Step 1: PhyNQ}}\\
    \For{each subject $i \in \mathcal{I}_{\text{train}}$}{
        Estimate per‑channel $1/f$ spectral slopes $\{\alpha_c\}$ from $\mathbf{X}_i$\;
        Use Otsu to obtain $\alpha_{\text{normal}}$ and $\alpha_{\text{anomaly}}$\;
        Compute channel‑wise noise $\nu_c$\;
        Subject‑level noise prior $\nu_i \gets \frac{1}{C}\sum_c \nu_c$\;
    }
    
    \tcp*[h]{\textbf{Step 2: DALR}} \\
    \For{each subject $i \in \mathcal{I}_{\text{train}}$}{
        Contamination ratio $\nu_i$\;
        Fit Isolation Forest on $\mathbf{Z}_i$, obtain anomaly scores $\tilde{\mathbf{s}}_i$\ and normal sample indices $\mathcal{C}_i$;
        
        GNB predicts soft pseudo‑labels $\tilde{\mathbf{y}}_{i,j}$ on $\{\mathbf{Z}^{\mathcal{C}},\mathbf{Y}^{O,\mathcal{C}}\}$ \;
        Compute score $conf_{i,j}$ and $\psi_{i,j}$ \;
        Use Otsu to split into $\mathcal{D}^{clean}$ and $\mathcal{D}^{noisy}$\;
        \For{each noisy sample $j$}{
            $\hat{\mathbf{y}}_{i,j} = conf_{i,j} \cdot \mathbf{y}_{i,j}^O + (1-conf_{i,j}) \cdot \tilde{\mathbf{y}}_{i,j}$\;
        }
        Keep clean sample labels as $\mathbf{y}_{i,j}^O$\;
        $\mathbf{Y}'_i \gets$ all refined labels of subject $i$\;
    }
    $\mathbf{Y}' \gets \bigcup_{i\in\mathcal{I}_{\text{train}}} \mathbf{Y}'_i$\;
}
\Return $\mathbf{Y}'$;
\end{algorithm}

%% file: 4_Implementation_Details.tex
\section{Implementation Details}
\subsection{Dataset}
We conducted experiments on three public datasets: DEAP~\cite{2011deap}, SEED~\cite{zheng2015SEED}, and SEED-IV~\cite{zheng2018SEEDIV}. All of them provide multi-channel EEG recordings from subjects during emotional stimulation. Table~\ref{tab:dataset_summary} summarizes their key statistics.

All three datasets were collected with ethical approval
from their respective institutional review boards, and informed consent was obtained from all participants.

\paragraph{DEAP} contains 32-channel EEG recorded from 32 subjects, with each one watching 40 one-minute video clips. Participants rated affective responses on 9-point scales for Arousal, Valence, Dominance, and Liking. Following \cite{2025MoCERNet,zhu2026bimoe}, we binarize ratings at threshold 5, classifying scores $>5$ as ``high'' and $\le 5$ as ``low''. We focus on Arousal as it is the most widely used dimension.

\paragraph{SEED} comprises 62 channels from 15 subjects, each participating in three sessions spaced approximately one week apart. In each session, subjects watched 15 carefully selected film clips intended to elicit positive, neutral, or negative emotions. The ground-truth labels are obtained from each subject's self-report immediately after viewing.

\paragraph{SEED-IV} extends the SEED paradigm to four emotion categories: happy, sad, neutral, and fear. It shares the same 15 subjects, 62-channel EEG, and three-session structure, but each session includes 24 film clips (six per category). Labels are also obtained via self-reports, providing a multi-class emotion recognition benchmark.

\input{tabs/table_DATA}
\input{tabs/table_deapA}
\input{tabs/table_seed}
\input{tabs/table_seediv}

\subsection{Data Preprocessing and Splitting}
To ensure reproducibility and fair comparison, all datasets are preprocessed and split under the LibEER benchmark~\cite{liu2025libeer}.
\paragraph{Preprocessing}
The preprocessing pipeline consists of bandpass filtering between 0.3 and 50 Hz, ocular artifact removal via PCA, extraction of differential entropy (DE) features across five frequency bands, i.e., [0.5, 4], [4, 8], [8, 14], [14, 30], and [30, 50] Hz, smoothed by a Linear Dynamic System (LDS), and segmentation into non-overlapping 1-second windows where each window inherits the trial-level label.

\paragraph{Splitting}
To rigorously evaluate the cross-subject generalization capability of the proposed label denoising method, we adopt a strict LOSO-CV protocol. In this setting, each subject is iteratively held out as the unseen test set while the model is trained on all remaining subjects, with final performance obtained by averaging results across all folds. 

This subject-independent evaluation strategy is widely recognized as the gold standard for assessing cross-subject EEG decoding performance \cite{li2026LOSOsurvey}. This ensures that the evaluation faithfully reflects the model's ability to generalize to entirely unseen individuals.

\subsection{Baseline and Backbone Selection}
To evaluate the proposed PhyDA framework comprehensively, we compare it with a diverse set of baselines \cite{lidividemix2020, 2021Colearning, 2025real, hai2025DDSPR} in seven backbones \cite{zheng2014DBN,li2018BiDANN, tao2020ACRNN,gao2020CDCN,wang2022HSLT,ding2022TSception,yang2025NSALDGAT} that represent different denoising paradigms.
\paragraph{Baselines}
These baselines were chosen because they cover different strategies for handling label noise, including the unmodified baseline (no denoising), sample selection \cite{2021Colearning}, robust loss functions \cite{2025real}, and label correction \cite{lidividemix2020, hai2025DDSPR}, with the latest and tailored EEG denoising methods \cite{2025real, hai2025DDSPR} among them. This selection covers a range of different approaches, ensuring that our evaluation considers the main strategies employed by existing denoising solutions. This provides a comprehensive reference against which the proposed method can be assessed.

All methods operate on identical pre‑extracted features and strictly adhere to the LOSO-CV protocol. For DivideMix \cite{lidividemix2020} and Co‑learning \cite{2021Colearning}, we build upon the publicly available code of the original paper, adapting them to the EEG setting as follows: (1) we disable all image‑specific data augmentations (e.g., MixUp, MixMatch, temperature sharpening), which would otherwise generate out‑of‑distribution virtual samples in cross‑subject EEG scenarios; (2) we replace the original vision‑oriented backbone with a lightweight, unified MLP (identical across all baselines) to ensure that performance differences stem only from the denoising logic. For REAL \cite{2025real} and DDSPR \cite{hai2025DDSPR}, we reproduce their noise‑filtering module based on the paper’s description.

\paragraph{Backbones}
To ensure that our evaluation does not hinge on any single architectural design, we select seven backbone models from the LibEER framework \cite{liu2025libeer} to cover major architectural paradigms in EEG-based emotion recognition: CDCN \cite{gao2020CDCN} and TSception \cite{ding2022TSception} (CNN‑based), ACRNN \cite{tao2020ACRNN} and BiDANN \cite{li2018BiDANN} (RNN‑based), NSAL‑DGAT \cite{yang2025NSALDGAT} (GNN‑based), DBN \cite{zheng2014DBN} (DNN‑based), and HSLT \cite{wang2022HSLT} (Transformer‑based).

\subsection{Experiment Settings}
All experiments were conducted on the high-performance server equipped with NVIDIA RTX 3090 Ti GPU, using Python 3.10.18 and PyTorch 2.7.1 framework with CUDA 12.6 support.

For all baselines, we adopt the default hyperparameters recommended in the original publications as the starting point. A small number of task‑dependent parameters are tuned via a small-scale grid search, and others remain at their defaults.
All of the backbones used in our experiments adopt the default hyperparameters provided by the LibEER framework \cite{liu2025libeer}.
The hyperparameter settings are reported in Table \ref{tab:hyperparams}.

%% file: tabs/table_DATA.tex
\begin{table}[htbp]
\centering
\caption{Summary of the DEAP, SEED, and SEED-IV datasets.}
\label{tab:dataset_summary}
\begin{tabular}{lccc}
\toprule
\textbf{Dataset} & \textbf{Subjects} & \textbf{Channels} &  \textbf{Emotion labels} \\
\midrule
DEAP    & 32 & 32 &  High, Low \\
SEED    & 15 & 62 & Positive, Neutral, Negative \\
SEED-IV & 15 & 62  & Happy, Sad, Neutral, Fear \\
\bottomrule
\end{tabular}
\end{table}

%% file: tabs/table_deapA.tex
\begin{table*}[t]
\caption{Performance Comparison on DEAP-A Dataset (\textit{Acc} and \textit{wF1}) for LOSO-CV Experiment}
\setlength{\tabcolsep}{2.5mm}
\resizebox{1.0\linewidth}{!}{
\label{tab:deap-a}
\centering
\small
\begin{tabular}{cccccccccc}
\toprule
\multirow{2}{*}{\textbf{Method}} & \multirow{2}{*}{\textbf{Metric}} & \multicolumn{2}{c}{CNN-based} & \multicolumn{2}{c}{RNN-based} & GNN-based & DNN-based & Transformer-based & Avg.\\
\cmidrule(lr){3-4} \cmidrule(lr){5-6} \cmidrule(lr){7-7} \cmidrule(l){8-8} \cmidrule(l){9-9}
 &  & CDCN & TSception & ACRNN & BiDANN & NSAL-DGAT & DBN & HSLT & Gain\\
\midrule
Baseline & \textit{Acc} & 62.13±9.06 & 63.33±10.84 & 61.42±11.79 & 63.81±11.03 & 63.17±9.39 & 66.18±9.23 & 55.24±10.12 & -- \\
 & \textit{wF1} & 59.29±9.96 & 59.03±12.51 & 55.49±12.25 & 57.39±12.11 & 54.52±11.70 & 57.58±10.83 & 52.47±11.79 & -- \\
\addlinespace

DivideMix & \textit{Acc} & 60.63±9.81 & 61.51±11.83 & 61.50±12.56 & 60.75±12.29 & 61.97±10.23 & 66.98±9.94 & 57.06±12.58 & -0.70 \\
 & \textit{wF1} & 59.35±10.89 & 57.37±12.46 & 55.10±11.24 & 57.50±12.71 & 55.14±12.63 & 57.60±12.26 & 52.44±12.60 & -0.18 \\
\addlinespace

Co-learning & \textit{Acc} & 62.78±10.41 & 61.72±12.07 & 58.07±13.98 & 64.39±9.14 & 64.43±10.21 & 62.57±10.19 & 56.12±9.82 & -0.74 \\
 & \textit{wF1} & 58.09±11.56 & 54.60±12.45 & 54.98±14.07 & 58.27±10.50 & 56.46±11.15 & 54.66±11.78 & 51.41±10.38 & -1.04 \\
\addlinespace

DDSPR & \textit{Acc} & 63.59±12.54 & 64.94±12.63 & 63.75±12.16 & 64.95±9.90 & 64.84±11.80 & 66.54±12.47 & 57.83±13.80 & +1.59 \\
 & \textit{wF1} & 59.76±12.38 & 57.79±11.09 & 57.32±10.58 & 62.14±9.50 & 55.09±11.38 & 57.98±11.63 & 50.91±14.25 & +0.75 \\
\addlinespace

REAL & \textit{Acc} & 64.02±9.49 & 65.78±12.31 & 62.97±13.50 & 65.93±11.32 & 63.80±9.69 & \cellcolor{gray!40}67.98±9.97 & \cellcolor{gray!40}61.79±11.73 & +2.43 \\
 & \textit{wF1} & 60.62±10.24 & \cellcolor{gray!40}60.91±10.43 & 57.08±12.63 & 60.60±12.28 & 57.33±10.17 & 59.41±11.20 & \cellcolor{gray!40}55.47±12.61 & +2.24 \\
\addlinespace

PhyDA (Ours) & \textit{Acc} & \cellcolor{gray!40}64.25±8.75 & \cellcolor{gray!40}66.27±9.00 & \cellcolor{gray!40}64.29±11.57 & \cellcolor{gray!40}67.25±9.74 & \cellcolor{gray!40}64.86±9.41 & 67.74±10.67 & 59.97±12.18 & \cellcolor{gray!40}+2.76 \\
 & \textit{wF1} & \cellcolor{gray!40}61.03±9.57 & 60.41±10.50 & \cellcolor{gray!40}58.32±12.35 & \cellcolor{gray!40}63.26±9.82 & \cellcolor{gray!40}58.77±10.01 & \cellcolor{gray!40}59.59±10.54 & 54.06±13.87 & \cellcolor{gray!40}+2.81 \\
\bottomrule 
\end{tabular}
}
\end{table*}

%% file: tabs/table_seed.tex
\begin{table*}[t]
\caption{Performance Comparison on SEED Dataset (\textit{Acc} and \textit{wF1}) for LOSO-CV Experiment}
\setlength{\tabcolsep}{2.5mm}
\resizebox{1.0\linewidth}{!}{
\label{tab:seed}
\centering
\small
\begin{tabular}{cccccccccc}
\toprule
\multirow{2}{*}{\textbf{Method}} & \multirow{2}{*}{\textbf{Metric}} & \multicolumn{2}{c}{CNN-based} & \multicolumn{2}{c}{RNN-based} & GNN-based & DNN-based & Transformer-based & Avg.\\
\cmidrule(lr){3-4} \cmidrule(lr){5-6} \cmidrule(lr){7-7} \cmidrule(l){8-8} \cmidrule(l){9-9}
 &  & CDCN & TSception & ACRNN & BiDANN & NSAL-DGAT & DBN & HSLT & Gain\\
\midrule
\multirow{2}{*}{Baseline} & \textit{Acc} & 62.65±7.62 & 65.95±9.92 & 64.17±11.43 & 63.91±10.25 & 70.52±9.71 & 58.39±11.83 & 61.48±9.08 & -- \\
 & \textit{wF1} & 59.27±8.33 & 60.86±12.06 & 60.55±14.51 & 60.51±11.19 & 69.83±8.40 & 51.33±14.75 & 57.69±10.84 & -- \\
\addlinespace
\multirow{2}{*}{DivideMix\cite{lidividemix2020}} & \textit{Acc} & 59.77±12.85 & 60.13±12.17 & 61.47±12.60 & 60.56±11.93 & 66.12±12.48 & 58.07±11.50 & 61.20±10.88 & -2.82 \\
 & \textit{wF1} & 57.01±11.17 & 55.42±11.82 & 51.92±12.66 & 54.53±13.08 & 65.74±13.39 & 54.96±13.46 & 56.76±10.10 & -3.39 \\
\addlinespace
\multirow{2}{*}{Co-learning\cite{2021Colearning}} & \textit{Acc} & 62.54±10.37 & 62.39±10.65 & 64.50±14.04 & 65.50±12.42 & 69.39±9.87 & 58.32±12.07 & 58.82±12.53 & -0.80 \\
 & \textit{wF1} & 58.78±13.25 & 57.26±10.82 & 60.57±14.25 & 60.94±13.41 & 66.56±10.64 & 52.54±13.86 & 54.40±14.97 & -1.28 \\
\addlinespace
\multirow{2}{*}{DDSPR\cite{hai2025DDSPR}} & \textit{Acc} & \cellcolor{gray!40}66.39±7.83 & 66.67±9.75 & 66.73±10.65 & \cellcolor{gray!40}67.38±9.52 & 69.15±10.95 & 57.73±10.93 & 61.31±10.53 & +1.18 \\
 & \textit{wF1} & \cellcolor{gray!40}62.97±8.41 & 62.59±12.04 & 61.32±11.42 & 63.15±11.64 & 68.06±10.36 & 52.70±12.95 & 59.76±9.60 & +1.50 \\
\addlinespace
\multirow{2}{*}{REAL\cite{2025real}} & \textit{Acc} & 64.82±9.72 & 66.98±12.08 & 66.18±11.69 & 65.50±10.16 & 70.96±9.65 & 59.35±9.96 & 63.03±8.92 & +1.39 \\
 & \textit{wF1} & 61.35±8.28 & 64.41±12.22 & 60.42±13.74 & 62.88±10.94 & 70.72±9.25 & 53.98±12.51 & \cellcolor{gray!40}60.67±9.47 & +2.06 \\
\addlinespace
\multirow{2}{*}{PhyDA (Ours)} & \textit{Acc} & 66.19±6.95 & \cellcolor{gray!40}68.20±9.39 & \cellcolor{gray!40}67.51±9.72 & 66.16±8.15 & \cellcolor{gray!40}72.86±8.33 & \cellcolor{gray!40}61.32±10.25 & \cellcolor{gray!40}63.45±7.41 & \cellcolor{gray!40}+2.66 \\
 & \textit{wF1} & 62.45±7.95 & \cellcolor{gray!40}64.95±10.56 & \cellcolor{gray!40}62.90±10.10 & \cellcolor{gray!40}63.27±10.32 & \cellcolor{gray!40}71.85±8.83 & \cellcolor{gray!40}55.25±11.01 & 59.85±10.43 & \cellcolor{gray!40}+2.93 \\
\bottomrule 
\end{tabular}
}
\end{table*}

%% file: tabs/table_seediv.tex
\begin{table*}[t]
\caption{Performance Comparison on seed-iv Dataset (\textit{Acc} and \textit{wF1}) for LOSO-CV Experiment}
\setlength{\tabcolsep}{2.5mm}
\resizebox{1.0\linewidth}{!}{
\label{tab:seediv}
\centering
\small
\begin{tabular}{cccccccccc}
\toprule
\multirow{2}{*}{\textbf{Method}} & \multirow{2}{*}{\textbf{Metric}} & \multicolumn{2}{c}{CNN-based} & \multicolumn{2}{c}{RNN-based} & GNN-based & DNN-based & Transformer-based & Avg.\\
\cmidrule(lr){3-4} \cmidrule(lr){5-6} \cmidrule(lr){7-7} \cmidrule(l){8-8} \cmidrule(l){9-9}
 &  & CDCN & TSception & ACRNN & BiDANN & NSAL-DGAT & DBN & HSLT & Gain\\
\midrule
\multirow{2}{*}{Baseline} & \textit{Acc} & 44.68±7.85 & 49.96±8.19 & 43.96±8.09 & 49.65±9.53 & 48.11±6.22 & 48.25±7.59 & 41.68±7.57 & -- \\
 & \textit{wF1} & 37.57±10.01 & 41.59±10.14 & 34.37±9.27 & 44.27±12.01 & 38.11±8.94 & 40.38±10.33 & 32.49±10.85 & -- \\
\addlinespace
\multirow{2}{*}{DivideMix\cite{lidividemix2020}} & \textit{Acc} & 44.31±9.69 & 46.03±9.62 & 44.21±9.73 & 48.19±9.86 & 47.96±7.34 & 47.49±9.75 & 42.40±8.22 & -0.81 \\
 & \textit{wF1} & 38.13±11.51 & 39.24±12.08 & 34.34±9.58 & 40.12±11.66 & 38.55±8.04 & 39.44±10.69 & 33.18±10.56 & -0.83 \\
\addlinespace
\multirow{2}{*}{Co-learning\cite{2021Colearning}} & \textit{Acc} & 45.73±8.08 & 48.54±9.86 & 43.50±10.22 & 49.74±11.23 & 49.81±9.10 & 48.33±8.42 & 42.57±8.60 & +0.28 \\
 & \textit{wF1} & 38.66±9.47 & 40.70±11.04 & 33.39±10.15 & 44.79±12.26 & 38.91±9.30 & 41.04±10.86 & 33.10±10.81 & +0.26 \\
\addlinespace
\multirow{2}{*}{DDSPR\cite{hai2025DDSPR}} & \textit{Acc} & \cellcolor{gray!40}47.72±7.90 & 51.89±9.65 & 47.55±8.63 & 50.46±9.77 & 50.96±6.59 & 48.76±7.27 & \cellcolor{gray!40}44.81±7.33 & +2.27 \\
 & \textit{wF1} & \cellcolor{gray!40}39.40±8.76 & 42.28±10.03 & 36.96±10.47 & 45.56±10.62 & 39.85±8.59 & 41.33±9.34 & \cellcolor{gray!40}33.56±8.62 & +1.45 \\
\addlinespace
\multirow{2}{*}{REAL\cite{2025real}} & \textit{Acc} & 45.87±8.36 & 50.63±9.17 & 47.39±9.60 & 52.27±8.81 & 51.03±8.05 & 49.79±7.25 & 44.41±7.66 & +2.16 \\
 & \textit{wF1} & 39.04±9.24 & 40.63±12.19 & 37.26±10.04 & 45.12±11.49 & 40.38±9.92 & 42.03±10.51 & 32.97±9.08 & +1.24 \\
\addlinespace
\multirow{2}{*}{PhyDA (Ours)} & \textit{Acc} & 47.36±8.67 & \cellcolor{gray!40}52.78±9.53 & \cellcolor{gray!40}49.70±9.18 & \cellcolor{gray!40}52.90±7.74 & \cellcolor{gray!40}51.53±7.62 & \cellcolor{gray!40}50.50±7.88 & 44.75±7.31 & \cellcolor{gray!40}+3.32 \\
 & \textit{wF1} & 39.20±9.74 & \cellcolor{gray!40}42.34±9.76 & \cellcolor{gray!40}38.94±11.18 & \cellcolor{gray!40}45.84±10.53 & \cellcolor{gray!40}40.74±9.35 & \cellcolor{gray!40}42.94±9.68 & 33.12±8.82 & \cellcolor{gray!40}+2.05 \\
\bottomrule
\end{tabular}
}
\end{table*}

%% file: 5_experiment.tex
\input{tabs/table_hyper}
\section{Experiments}
\begin{figure*}[!ht]
\centerline{\includegraphics[scale=0.90]{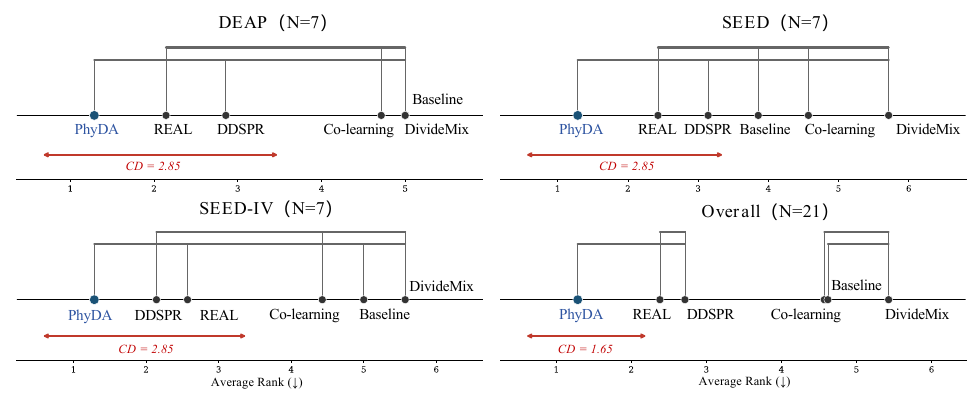}}
    \caption{Critical difference diagrams for the three datasets and the overall 21-task comparison at $\alpha=0.05$ (Nemenyi test). PhyDA achieves the lowest (best) average rank in all four panels.
    }
    \vspace{-5pt}
    \label{fig:cd}
\end{figure*}
\input{tabs/Ablation}
\subsection{Comparative Experiments}
To evaluate PhyDA comprehensively, we conducted experiments on three public datasets (DEAP, SEED, and SEED-IV) using seven backbone architectures and five competing methods. All experiments were performed using publicly available benchmark \cite{liu2025libeer} and under strict LOSO-CV protocols. The detailed results are reported in Tables~\ref{tab:deap-a},~\ref{tab:seed}, and~\ref{tab:seediv} for DEAP, SEED, and SEED‑IV, respectively. These three datasets correspond to binary, three-class, and four-class classification tasks, with the difficulty of classification gradually increasing. For each method and backbone, we report the average classification accuracy (\textit{Acc}), weighted F1-score (\textit{wF1}), the corresponding standard deviation (\textit{±std}), and the average gain (\textit{Avg.\ Gain}) relative to the baseline without label enhancement. The best result in each column is highlighted with a gray background.

The results demonstrate that PhyDA consistently improves performance across all backbones and achieves the highest \textit{Avg.\ Gain} on all three datasets, with improvements of 2.76\%, 2.66\%, and 3.32\% over the respective baselines on DEAP, SEED, and SEED‑IV. This consistent superiority stems from the unique design of PhyDA: (1) the PhyNQ module exploits the $1/f^\alpha$ power-law decay, a fundamental physiological characteristic, to provide a neurophysiologically interpretable prior and a preliminary data selection; (2) all thresholds are computed in a subject‑independent manner, giving the algorithm strong adaptability and cross‑subject generalization; (3) PhyDA retains all samples without discarding any data, preserving the feature information of noisy instances. 

Additionally, among the competing methods, DDSPR \cite{hai2025DDSPR} and REAL \cite{2025real} consistently outperform DivideMix \cite{lidividemix2020} and Co‑learning \cite{2021Colearning}. This can be attributed to REAL explicitly modeling EEG’s temporal variability via local feature aggregation and iterative confidence filtering, while DDSPR addresses cross‑subject domain shifts through dynamic domain selection and pseudo‑label refinement. In contrast, Co‑learning lacks explicit modeling of the interdependencies among EEG segments, rendering it ineffective in handling intra‑subject non‑stationarity. DivideMix, on the other hand, relies on noise assumptions (e.g., symmetric or random label noise) that are ill‑suited for EEG data and fails to account for domain distribution shifts. Although DDSPR and REAL achieve notable performance gains, they do so by discarding low-confidence or noisy samples, which is a wasteful strategy given the limited size of EEG datasets.

To formally assess the ranking stability, we conducted a Friedman test \cite{pereira2015} on each dataset and on the pooled 21 dataset-backbone combinations. In all four cases the null hypothesis of equal performance was rejected ($p<0.001$). Post-hoc Nemenyi analysis over the 21 tasks (CD = 1.645) confirmed that PhyDA was significantly higher than Baseline, DivideMix, and Co-learning, while its difference from DDSPR and REAL was not statistically separable under the current task count. The critical difference diagrams for all four settings are shown in Fig.~\ref{fig:cd}.

\begin{figure*}[!ht]
\centerline{\includegraphics[scale=1.3]{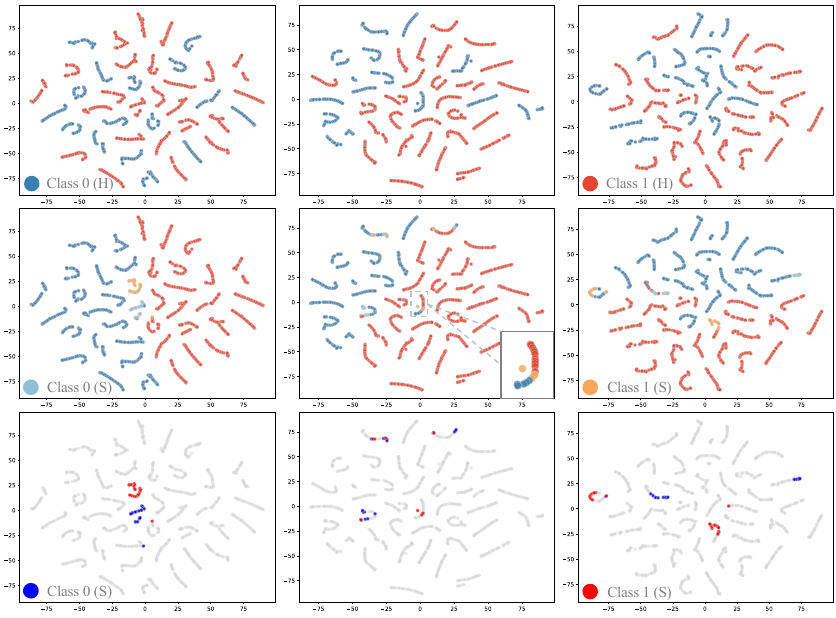}}
    \caption{Visualization of 2D t-SNE on DEAP. Each column corresponds to the data from a randomly selected subject. \textbf{Top row}: Visualization of the original label. \textbf{Middle row}: The enhanced labels show hard labels (H) in darker colors and soft labels (S) in lighter colors. \textbf{Bottom row}: Highlight the soft labels against a transparent background, it can be seen that they are concentrated at the classification boundaries.
    }
    \vspace{-5pt}
    \label{fig2}
\end{figure*}

\begin{figure}[!ht]
\centerline{\includegraphics[scale=0.8]{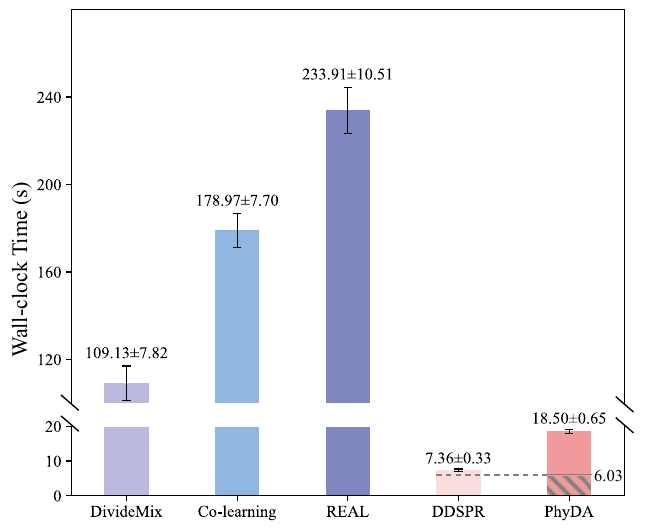}}
    \caption{Wall-clock Time Comparison on DEAP
    }
    \vspace{-5pt}
    \label{fig3}
\end{figure}

\subsection{Ablation Experiments}
To evaluate the contribution of each component in PhyDA, we design six variants and conduct ablation experiments across all three datasets. To ensure that the results are not tied to a single architectural paradigm, we select one representative backbone from each major category: BiDANN (RNN-based) on DEAP, TSception (CNN-based) on SEED, and NSAL-DGAT (GNN-based) on SEED-IV, all under a strict LOSO-CV setting. Specifically, we replace the contamination ratio estimate $\textbf{v}'$ for each subject based on PhyNQ with the default IForest parameter (contamination=0.1), to obtain the result of \textit{w/o PhyNQ}, which tests the necessity of the physiological prior. To further verify that the $1/f^\alpha$ spectral slope is uniquely valuable, we additionally design \textit{w/ Var Proxy}: it replaces the spectral slope $\alpha_c$ with the log‑transformed temporal variance of the filtered EEG signal, thereby providing a simple, non‑physiological contamination estimator of equal complexity. For \textit{w/o IForest}, we bypass the anomaly detection step and directly use  $\textbf{v}'$ as the anomaly scores $\tilde{s}$, isolating the contribution of IForest. To assess the role of GNB, we replace it with a two‑layer MLP trained on the same clean subset for pseudo‑label generation (\textit{w/o GNB}). Finally, we replace the automatic Otsu thresholding in both PhyNQ and DALR with fixed manual split ratios of 30\%, 50\%, and 70\%, denoted as \textit{w/o $\tau^{(1)}$} and \textit{w/o $\tau^{(2)}$} respectively, to examine the importance of adaptive threshold selection.

As shown in Table~\ref{tab:ablation}, across all three datasets, removing the PhyNQ module consistently results in a significant performance decrease, confirming the necessity of a subject-specific, physiologically based contamination estimate. Notably, \textit{w/ Var Proxy} underperforms the full PhyDA model across datasets, confirming that the $1/f^\alpha$ spectral slope captures noise-related information beyond what a simple statistical measure can provide. The result is also degraded without IForest, which verifies that data-driven anomaly scoring provides complementary information beyond the physiological prior. Using an MLP instead of a GNB reduces performance across datasets, because in scenarios with small samples, the MLP is prone to overfitting, which results in unreliable pseudo-labels. Additionally, the MLP requires extra training costs and manual parameters adjustment. Furthermore, we found that, when using a fixed threshold, the results were consistently inferior to those obtained using subject-adaptive partitioning based on Otsu. This is because the fixed threshold disregarded the wide differences in the subjects' data. Overall, the ablation confirms that every module in PhyDA contributes positively to the final performance across datasets and backbone architectures, and the framework achieves its best results only when all components operate jointly.

\subsection{Complexity Comparison}
We evaluate the computational efficiency of PhyDA against four representative label enhancement methods. Since PhyDA does not rely on trainable neural network modules, conventional metrics such as FLOPs and trainable parameter counts are not applicable. We therefore adopt wall‑clock time (Times) as the primary and fairest comparison metric. The average processing times and standard deviations over five runs are reported in Fig.~\ref{fig3}. As shown, PhyDA requires $18.50$ s to process all 32 subjects, which is higher than DDSPR's $7.36$ s but substantially faster than DivideMix ($109.13$ s), Co‑learning ($178.97$ s), and REAL ($233.91$ s).

Since DivideMix, Co-learning, and REAL require additional neural networks to be trained, the computing time is much longer than PhDA and DDSPR. Compared with DDSPR, the extra time required for PhDA originates from its PhNQ stage. This per-subject fitting introduces additional overhead, accounting for approximately $67\%$ ($12.47$ s) of the total. But PhNQ operates purely on a per-subject basis, and its output is independent of the splitting. Therefore, throughout the entire LOSO-CV procedure, PhNQ needs to be executed only once. The DALR component takes only $6.03$ s, which is very close to the total processing time of DDSPR. Given PhyDA's consistent superiority over DDSPR in terms of overall performance, the additional computational cost is considered acceptable.

\subsection{Visualization Analysis}
To intuitively assess the effectiveness of PhDA, we randomly selected three subjects from the DEAP dataset and visualized their features in 2-dimensional (2D) t-SNE \cite{maaten2008tsne}.  
The results are shown in Fig.~\ref{fig2}, where each column corresponds to one subject, and each row illustrates a distinct labeling scheme. The top row depicts the original labels.  
The \textbf{middle row} shows the enhanced labels produced by PhDA. The light colors represent soft labels (S) with low confidence levels, and the darker colors represent hard labels (H). 
A clear comparison between the \textbf{top row} and middle row reveals that PhDA yields noticeably sharper decision boundaries.  
The \textbf{bottom row} explicitly highlights the samples with soft labels. We can further discover that these samples concentrate along the decision boundary, precisely where the original labels are most uncertain. 

\begin{figure}[!ht]
\centerline{\includegraphics[scale=0.75]{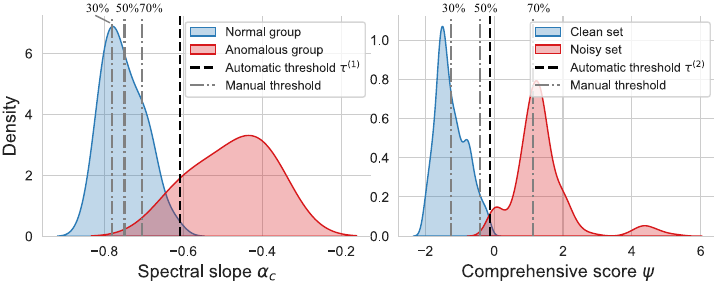}}
    \caption{Visual comparison of automatic threshold and manual threshold (including the 30\%, 50\% and 70\% percentiles). \textbf{Left}: Spectral slope division. \textbf{Right}: Comprehensive score division.
    }
    \vspace{-5pt}
    \label{fig4}
\end{figure}

We further examine the two Otsu-based splits by visualizing the distributions of the spectral slope $\alpha_c$ and the comprehensive score $\psi$ for the first subject in Fig.~\ref{fig4}. Both histograms display a clear bimodal shape, with the Otsu thresholds $\tau^{(1)}$ and $\tau^{(2)}$ (black dashed lines) automatically located near the valleys between the two sets. In contrast, fixed manual thresholds at the 30\%, 50\%, and 70\% percentiles (gray dashed lines) produce visibly unreasonable splits. This observation is consistent with the ablation results in Table~\ref{tab:ablation}, where the performance of fixed manual threshold variants is consistently lower than the adaptive Otsu strategy.

Overall, the PhDA can correct incorrect labels and soften the labels of difficult samples. This effectively reduces the impact of label noise and significantly improves the model's cross-subject generalization performance. Both PhyNQ and DALR use a reasonable adaptive partitioning strategy that adapts the thresholds for each subject without manual intervention.

%% file: tabs/table_hyper.tex
\begin{table}[htbp]
\centering
\caption{Hyperparameter settings.}
\label{tab:hyperparams}
\begin{tabular}{c l l}
\toprule
\textbf{Method} & \textbf{Hyperparameter} & \textbf{Value} \\
\midrule
\multirow{2}{*}{DivideMix}           & num\_epochs & 25 \\
                                     & initial noise\_rate & 0.5 \\
\midrule
\multirow{2}{*}{Co‑learning}         & epochs & 50 \\
                                     & forget\_rate & 0.3 \\
                                     
\midrule
\multirow{2}{*}{DDSPR} & Number of similar subjects & 7 \\
                                     & confidence threshold & 0.5 \\

\midrule
\multirow{4}{*}{REAL}                & epochs & 150 \\
                                     & forget\_rate & 0.25 \\
                                     & num\_gradual & 10 \\
                                     & conf\_threshold & 0.8 \\
\midrule
\multirow{4}{2.5cm}{\centering General hyperparameters \\ in backbones}
                                     & learning rate & 0.001 \\
                                     & batch\_size & 128 \\
                                     & epochs & 40 \\
                                     & random seed & 1 \\
\bottomrule
\end{tabular}
\end{table}

%% file: tabs/Ablation.tex
\begin{table*}[t]
\centering
\caption{Ablation results across three datasets with representative backbones covering RNN, CNN, and GNN architectures.}
\setlength{\tabcolsep}{2.5mm}

\label{tab:ablation}
\resizebox{1.0\linewidth}{!}{
\begin{tabular}{lcc|ccccccc}
\toprule
\multirow{2}{*}{Dataset (Backbone)} & \multirow{2}{*}{Metric} & \multirow{2}{*}{PhyDA} & w/o & w/ Var & w/o & w/o & w/o $\tau^{(1)}$ & w/o $\tau^{(2)}$ \\
& & & PhyNQ & Proxy & IForest & GNB & (30\% / 50\% / 70\%) & (30\% / 50\% / 70\%) \\
\midrule
\multirow{2}{*}{DEAP (BiDANN)} & \textit{Acc} & \cellcolor{gray!40}67.25 & 65.80 & 65.12 & 64.87 & 64.95 & 65.10 / 66.67 / 66.26 & 65.74 / 66.31 / 66.92 \\
& \textit{wF1} & \cellcolor{gray!40}63.26 & 61.37 & 60.45 & 60.14 & 60.32 & 60.08 / 62.81 / 61.98 & 61.61 / 62.78 / 62.49 \\
\midrule
\multirow{2}{*}{SEED (TSception)} & \textit{Acc} & \cellcolor{gray!40}68.20 & 66.78 & 66.13 & 65.89 & 65.94 & 66.06 / 67.64 / 67.21 & 66.67 / 67.71 / 67.38  \\
& \textit{wF1} & \cellcolor{gray!40}64.95 & 63.12 & 62.35 & 61.74 & 62.27 & 62.10 / 63.84 / 63.32 & 62.75 / 63.95/ 63.23  \\
\midrule
\multirow{2}{*}{SEED-IV (NSAL-DGAT)} & \textit{Acc} & \cellcolor{gray!40}51.53 & 50.54 & 50.62 & 49.87 & 49.95 & 50.15 / 51.22 / 50.92 & 50.56 / 50.90 / 51.21 \\
& \textit{wF1} & \cellcolor{gray!40}40.74 & 39.66 & 39.74 & 39.34 & 39.10 & 39.39 / 40.16 / 39.83 & 39.62 / 40.18 / 40.33 \\
\bottomrule

\end{tabular}
}
\end{table*}

%% file: 6_conclusion.tex
\section{Conclusion}

We presented PhyDA, a tuning-free framework that unifies physiological modeling and data‑driven label enhancement. This improves the model's generalization performance in cross-subject scenarios for emotion recognition. PhyDA consists of two complementary modules: PhyNQ leverages the $1/f^\alpha$ spectral slope, a robust neurophysiological regularity, to predict noise ratios $\nu$ of each subject in a subject‑independent manner; DALR, a training‑free pipeline, adopts $\nu$ as the contamination prior for isolation forest anomaly detection, and combined with pseudo‑labels from a Gaussian Naive Bayes classifier, forms a composite scoring strategy to adaptively refine labels of noisy samples. Extensive experiments on three public datasets across seven backbones demonstrate that PhyDA achieves state‑of‑the‑art performance improvements over multiple strong baselines under a strict LOSO-CV setting. Visualization analyses further validate the effectiveness of the proposed approach.

{Two limitations of the current work should be mentioned. First, PhyNQ relies on the $1/f^\alpha$ spectral slope, which may be sensitive to differences in acquisition hardware and reference electrode placement. Second, the validation is currently limited to emotion recognition tasks.}

Future work may extend PhyDA to other EEG paradigms, such as motor imagery or mental workload assessment, where label noise is similarly prevalent, and explore the combination of the $1/f^\alpha$ prior with other physiological indicators for multimodal noise quantification. {PhyDA also holds promise for clinical BCI applications such as depression assessment and consciousness disorder detection, where subjective or unavailable behavioral labels make physiologically-based label refinement particularly valuable.}